\documentclass[12pt,preprint,epsf]{aastex}

\newcommand{\dmm}{\mbox{$\Delta$m$_{15}(B)$}}

\newcommand{\bvri}{\protect\hbox{$BV\!RI$} }

\newcommand{\yjhk}{\protect\hbox{$Y\!JHK_s$} }

\shorttitle{The Type~Ia Supernova 2005df}
\shortauthors{Krisciunas et al.} 

\begin{document}
\received{26 November 2017}

\title{Optical and Infrared Photometry of SN 2005df\altaffilmark{1}}
\author{
Kevin Krisciunas,\altaffilmark{2}
Nicholas B. Suntzeff,\altaffilmark{2}
Juan Espinoza,\altaffilmark{3} 
David Gonzalez,\altaffilmark{3} 
Alberto Miranda,\altaffilmark{3} 
and Pedro Sanhueza\altaffilmark{4} 
}
\altaffiltext{1}{Based in part on observations taken at the Cerro Tololo
Inter-American Observatory, National Optical Astronomy Observatory, 
which is operated by the Association of Universities for Research in 
Astronomy, Inc. (AURA) under cooperative agreement with the National 
Science Foundation.}
\altaffiltext{2}{George P. and Cynthia Woods Mitchell Institute for
Fundamental Physics and Astronomy, Texas A. \& M. University, Department of Physics,
  4242 TAMU, College Station, TX 77843;  {krisciunas@physics.tamu.edu} }
\altaffiltext{3}{Cerro Tololo Inter-American Observatory, Casilla 603,
  La Serena, Chile}
\altaffiltext{4}{Oficina de Protecci\'{o}n de la Calidad del Cielo del Norte
de Chile (OPCC), 1606 Cisternas, La Serena, Chile}


\begin{abstract}

We present optical (\bvri\hspace{-1.0mm}) and near-infrared
(\yjhk\hspace{-1.0mm}) photometry of the normal Type Ia supernova 2005df,
obtained with the CTIO 1.3-m and 0.9-m telescopes.
The $B$- and $V$-band photometry, S-corrected to the filter prescriptions
of \citet{Bes90}, matches the corresponding photometry from the ANU published
by \citet{Mil_etal10}.  The $R$-band photometry from CTIO and ANU matches
well without any corrections.  A combination of $V$-band and near-IR
photometry shows that SN~2005df is unreddened in its host galaxy.  Spectropolarimetry
of this supernova was obtained with the VLT, and the distance to the host
galaxy is being determined from observations of Cepheids using the Hubble
Space Telescope.  
\end{abstract}


SN~2005df was discovered visually by R. Evans on 2005 August 4.625 UT
some 15$^{\prime \prime}$ east and 40$^{\prime \prime}$ north of the nucleus of 
NGC 1559 \citep{Eva05}, the heliocentric radial velocity of which is
1304 km s$^{-1}$ \citep{Kor_etal04}. The supernova was located at RA = 04$^h$ 17$^m$
37.\hspace{-1.3mm}$^s$85, DEC =$-$62$^{\rm o}$ 46$^{\prime}$ 09$\farcs$5
(J2000).  See Fig. \ref{finder} for a finder chart. SN~2005df was confirmed to be a Type~Ia
supernova by \citet{Sal05} from a spectrum of August 5.83 UT taken with the
Australian National University 2.3-m telescope at Siding Spring. 

All of our photometry (except on one night) was taken with the CTIO 1.3-m 
telescope and the optical/IR imager ANDICAM. ANDICAM contains standard 
Johnson $UBV$ filters, Kron-Cousins $R$ and $I$ filters and standard 
Caltech/CTIO $JHK_s$ filters. Read out in 2 $\times$ 2 binning mode, ANDICAM 
gives a plate scale on the 1.3-m telescope of 0$\farcs$369 px$^{-1}$ for 
optical imaging and 0$\farcs$274 px$^{-1}$ for IR imaging. The optical field 
of view was 6$\farcm$3 by 6$\farcm$3, while the IR field of view was 
2$\farcm$34 by 2$\farcm$34.  ANDICAM also contains a 1.03 $\mu$m filter known
as $Y$.  This is an interesting photometric band for observing Type Ia supernovae,
because the second $Y$-band maximum of such an object is almost always brighter
than the first maximum.

Using optical standards of \citet{Lan92}, we calibrated 11 tertiary standards 
in the field of SN~2005df based on five nights of all-sky photometry with the 
CTIO 0.9-m telescope in November and December of 2006. The optical color 
terms for the CTIO 1.3-m telescope were determined from observations of the 
\citet{Lan92} field T Phe on two clear nights in September of 2005. These 
color terms were consistent with the mean color terms from four nights of 
February 2004. The $JHK_s$ magnitudes of two of the tertiary standards 
were calibrated on two clear nights in September 2005, using the near-IR 
standards P9104 and P9109 of \citet{Per_etal98}.  We adopted the $Y$-band 
magnitudes of these two Persson stars from \citet[][Appendix D]{Kri_etal17} 
to calibrate the $Y$-band magnitudes of the two tertiary standards. All of 
the CTIO 1.3-m photometry of the supernova was then calibrated using 
observations of the tertiary standards. A final night of $BVRI$ photometry of 
SN~2005df was obtained with the CTIO 0.9-m telescope about$\sim$100 days 
after the time of maximum light.

The optical and infrared photometry of the tertiary standards is given in 
Tables \ref{standards} and \ref{ir_stds}.  The optical and infrared 
photometry of SN~2005df is given in Tables \ref{opt_photom} and 
\ref{ir_photom}.

We present all but one night of our photometry of SN~2005df in Fig. 
\ref{bvri_yjhk}. The brightness and location of the SN did not require the use of 
host galaxy subtraction templates.  Our CTIO 1.3-m photometry is based on aperture 
photometry using a typical aperture of radius 10 px.  On nights of bad seeing a 
larger software aperture was used.

\citet{Mil_etal10} also present optical photometry of SN~2005df, from the 
Australian National University 1.0-m and 2.3-m telescopes.  By applying 
S-corrections to the CTIO $B$- and $V$-band data \citep{Kri_etal03} and 
putting our photometry on the system of \citet{Bes90}, we can effectively 
reconcile the photometry in these two bands.  Our $R$-band photometry 
requires no correction.  We cannot reconcile differences in the $I$-band 
photometry from different telescopes (up to 0.17 mag at $t$ = +15 days). The 
near-IR $JHK_s$ photometry was S-corrected to the photometric system of 
\citet{Per_etal98}.  The $BVJHK_s$ S-corrections are given in Table 
\ref{scorrs}.
   
From a fourth order polynomial fit the $B$-band data we find that maximum light 
occurred on JD 2,453,599.2 $\pm$ 0.3, at which time $B_{max}$ = 12.320 and $B-V$ = 
$-$0.083. $V_{max}$ = 12.396, $R_{max}$ = 12.375, and the near-IR maxima are best 
estimated from the photometry of our second night (JD 2,453,596.89). The maximum 
magnitudes have uncertainties of $\pm$0.02 mag. The decline rate \citep{Phi93}
\dmm = 1.12 mag, like many normal Type Ia SNe.  

We find that SN~2005df is bluer than SN~2001el by these amounts:
$\Delta (V-J)$ = $-0.420 \pm 0.032$,
$\Delta (V-H)$ = $-0.422 \pm 0.017$,
$\Delta (V-K_s)$ = $-0.500 \pm 0.039$ mag. For SN~2001el we
adopt a value of the total extinction of A$_V$ = 0.586 mag and host extinction of A$_V$ = 0.472
$\pm$ 0.025 mag \citep{Kri_etal07}. Using values of A$_{\lambda}$/A$_V$ for dust with R$_V$ = 2.15
in the host of SN~2001el, and R$_V$ = 3.1 dust in our Galaxy and in the host of SN~2004S
\citep{Kri_etal06}, the resulting host galaxy color excesses of SN~2005df are as follows:
E($V-J$) = $-0.041 \pm 0.032$,
E($V-H$) = 0.000 $\pm$ 0.017, and
E($V-K_s$) = $-0.053 \pm 0.039$.  This gives three estimates of the $V$-band
host galaxy extinction, namely $-$0.057 $\pm$
0.045, 0.000 $\pm$ 0.021, and $-$0.060 $\pm$ 0.044.  
The weighted mean is A$_V$ = $-0.018\; \pm$ 0.017 mag.  At face value we can
say that SN~2005df is unreddened in its host galaxy.   In that case we only need to 
correct the photometry of SN~2005df
for the Galactic reddening along the line of sight, namely E($B-V$) = 0.030 
$\pm$ 0.003 mag \citep{Sch_etal98}. Using conversion factors of \citet[][Table 
8]{Kri_etal06}, this translates to extinctions of A$_B$~=~0.122, A$_V$ = 0.093, 
A$_R$ = 0.077, A$_I$ = 0.056, A$_Y$ $\approx$ 0.040, A$_J$ = 0.026, 
A$_H$ = 0.017, and A$_{K_s}$ = 0.011 mag.

NGC 1559, the host of SN~2005df, is presently being observed by the Hubble 
Space Telescope to determine its distance using Cepheids (L. Macri, private 
communication). This will allow us to calibrate the absolute magnitudes at 
maximum light of the supernova. Spectrophotometry of SN~2005df has also been 
obtained at the VLT (A. Cikota, in preparation). Thus, SN~2005df is not just 
``one more nearby supernova''.  It will be a ``gold star'' object.

\vspace {1 cm}

\acknowledgments

The CTIO 0.9-m and 1.3-m telescopes are operated by the Small
and Moderate Aperture Research Telescope System (SMARTS) Consortium.

\newpage

\begin{deluxetable}{ccccc}
\tablewidth{0pc}
\tablecaption{Optical Tertiary Standards near SN 2005df\tablenotemark{a}\label{standards}}
\tablehead{   \colhead{Star ID\tablenotemark{b}} &
\colhead{$B$} & \colhead{$V$} & \colhead{$R$} & \colhead{$I$} }
\startdata
1  &      13.551 & 12.990 & 12.648 & 12.308 \\
2  &      13.697 & 13.051 & 12.674 & 12.315 \\
3  &      13.818 & 13.307 & 12.994 & 12.686 \\
4  &      15.573 & 14.961 & 14.600 & 14.240 \\
5  &      13.328 & 12.852 & 12.568 & 12.266 \\
6  &      13.671 & 13.056 & 12.693 & 12.335 \\
7  &      14.147 & 13.326 & 12.855 & 12.391 \\
8  &      14.678 & 13.878 & 13.433 & 13.022 \\
9  &      15.690 & 14.771 & 14.237 & 13.767 \\
10 &      16.403 & 15.515 & 15.009 & 14.562 \\
\enddata
\tablenotetext{a} {The mean errors of the mean of the photometry are $\pm$ 0.010 mag
or less.}
\tablenotetext{b} {The identifications are the same as in Fig. \ref{finder}.}
\end{deluxetable}

\begin{deluxetable}{ccccc}
\tablewidth{0pc}
\tablecaption{Infrared Tertiary Standards near SN~2005df\tablenotemark{a}\label{ir_stds}}
\tablehead{   \colhead{Star ID\tablenotemark{b}} &
\colhead{$Y$} & \colhead{$J$} & \colhead{$H$} & \colhead{$K_s$} }
\startdata
10 & 14.234 (0.015) & 13.905 (0.016) & 13.497 (0.013) & 13.358 (0.026) \\
11 & 12.524 (0.024) & 12.308 (0.013) & 12.040 (0.002) & 11.922 (0.016) \\
\enddata
\tablenotetext{a} {The numbers in parentheses are 1-$\sigma$ uncertainties (mean errors
of the mean).}
\tablenotetext{b} {The identifications are the same as those in Fig. \ref{finder}.}
\end{deluxetable}

\begin{deluxetable}{ccccc}
\tablewidth{0pc}
\tablecaption{\bvri Photometry of SN~2005df\tablenotemark{a}\label{opt_photom}}
\tablehead{   \colhead{JD$-$2,453,000} &
\colhead {$B$} & \colhead{$V$} & \colhead{$R$} & \colhead{$I$}  }
\startdata
591.87 & 12.909 (0.040) &  12.911 (0.021) &  12.846 (0.024) &  12.962 (0.026) \\
596.89 & 12.441 (0.028) &  12.475 (0.026) &  12.464 (0.032) &  12.707 (0.037) \\
600.86 & 12.415 (0.042) &  12.399 (0.031) &  12.425 (0.050) &  12.836 (0.028) \\
606.91 & 12.717 (0.028) &  12.538 (0.036) &  12.579 (0.024) &  13.099 (0.032) \\
614.87 & 13.530 (0.027) &  13.003 (0.027) &  13.018 (0.022) &  13.412 (0.028) \\
618.84 & 13.998 (0.029) &  13.234 (0.022) &  13.091 (0.024) &  13.287 (0.026) \\
624.80 & 14.597 (0.030) &  13.528 (0.024) &  13.250 (0.025) &  13.179 (0.029) \\
629.82 & 15.008 (0.040) &  13.817 (0.027) &  13.484 (0.030) &  13.222 (0.027) \\
632.76 & 15.173 (0.041) &  13.999 (0.018) &  13.695 (0.022) &  13.402 (0.027) \\
635.72 & 15.309 (0.039) &  14.156 (0.026) &  13.878 (0.024) &  13.600 (0.039) \\
700.79 & 16.401 (0.040) &  15.935 (0.030) &  16.012 (0.025) &  16.333 (0.040) \\
\enddata
\tablenotetext{a} {This photometry is in the photometric system of \citet{Lan92}.
The numbers in parentheses are 1-$\sigma$ uncertainties (mean
errors of the mean).  The first 10 nights' data were taken with the CTIO 1.3-m
telescope.  The final night's data were taken with the CTIO 0.9-m telescope.}
\end{deluxetable}

\begin{deluxetable}{ccccc}
\tablewidth{0pc}
\tablecaption{Near Infrared Photometry of SN~2005df\tablenotemark{a}\label{ir_photom}}
\tablehead{   \colhead{JD$-$2,453,000} & \colhead{$Y$} &
\colhead {$J$} & \colhead{$H$} & \colhead{$K_s$}  }
\startdata
591.87 &  13.076 (0.018) & 13.081 (0.054) & 13.234 (0.012) & 13.136 (0.044) \\
596.89 &  13.010 (0.018) & 12.888 (0.013) & 13.134 (0.010) & 12.938 (0.026) \\
600.86 &  13.357 (0.022) & 13.132 (0.016) & 13.335 (0.019) & 13.265 (0.036) \\
606.91 &  13.764 (0.028) & 13.865 (0.020) & 13.470 (0.022) & 13.426 (0.043) \\
614.87 &  13.724 (0.018) & 14.677 (0.017) & 13.388 (0.009) & 13.337 (0.018) \\
618.84 &  13.449 (0.018) & 14.583 (0.018) & 13.289 (0.010) & 13.263 (0.024) \\
624.80 &  13.149 (0.018) & 14.416 (0.021) & 13.213 (0.012) & 13.155 (0.024) \\
629.82 &  12.886 (0.017) & 14.147 (0.015) & 13.310 (0.009) & 13.342 (0.024) \\
632.76 &  12.919 (0.017) & 14.204 (0.015) & 13.449 (0.009) & 13.724 (0.039) \\
635.72 &  13.085 (0.021) & 14.516 (0.027) & 13.684 (0.016) & 13.941 (0.089) \\
\enddata
\tablenotetext{a} {This photometry is in the natural system of the CTIO 1.3-m 
telescope.  The numbers in parentheses are 1-$\sigma$ uncertainties (mean
errors of the mean).}
\end{deluxetable}

\begin{deluxetable}{cccccc}
\tablewidth{0pc}
\tablecaption{Photometric Corrections for SN 2005df\tablenotemark{a}\label{scorrs}}
\tablehead{  \colhead{JD$-$2,453,000} & \colhead{$\Delta B$} & \colhead{$\Delta V$} &
\colhead{$\Delta J$} & \colhead{$\Delta H$} & \colhead{$\Delta K_s$} }
\startdata
591.87 &  $-$0.044 &  $-$0.002 & \phs0.042 & \phs0.003 & \phs0.001 \\
596.89 &  $-$0.043 & \phs0.009 & \phs0.059 &  $-$0.008 & \phs0.018 \\
600.86 &  $-$0.044 & \phs0.017 & \phs0.068 &  $-$0.019 & \phs0.033 \\
606.91 &  $-$0.040 & \phs0.023 & \phs0.009 &  $-$0.043 & \phs0.054 \\
614.87 &  $-$0.019 & \phs0.026 &  $-$0.058 &  $-$0.046 & \phs0.022 \\
618.84 &  $-$0.006 & \phs0.028 &  $-$0.084 &  $-$0.060 & \phs0.008 \\
624.80 &  $-$0.027 & \phs0.032 &  $-$0.067 &  $-$0.070 &  $-$0.012 \\
629.82 &  $-$0.069 & \phs0.037 &  $-$0.075 &  $-$0.031 &  $-$0.011 \\
632.76 &  $-$0.069 & \phs0.039 &  $-$0.093 &  $-$0.009 &  $-$0.007 \\
635.72 &  $-$0.068 & \phs0.040 &  $-$0.107 & \phs0.007 &  $-$0.006 \\
\enddata
\tablenotetext{a} {The values are to be {\em added} to the corresponding
photometry in Tables \ref{opt_photom} and \ref{ir_photom}.  This places
the $BV$ photometry on the system of \citet{Bes90} and the $JHK_s$
photometry on the system of \citet{Per_etal98}.}
\end{deluxetable}

\clearpage

\figcaption[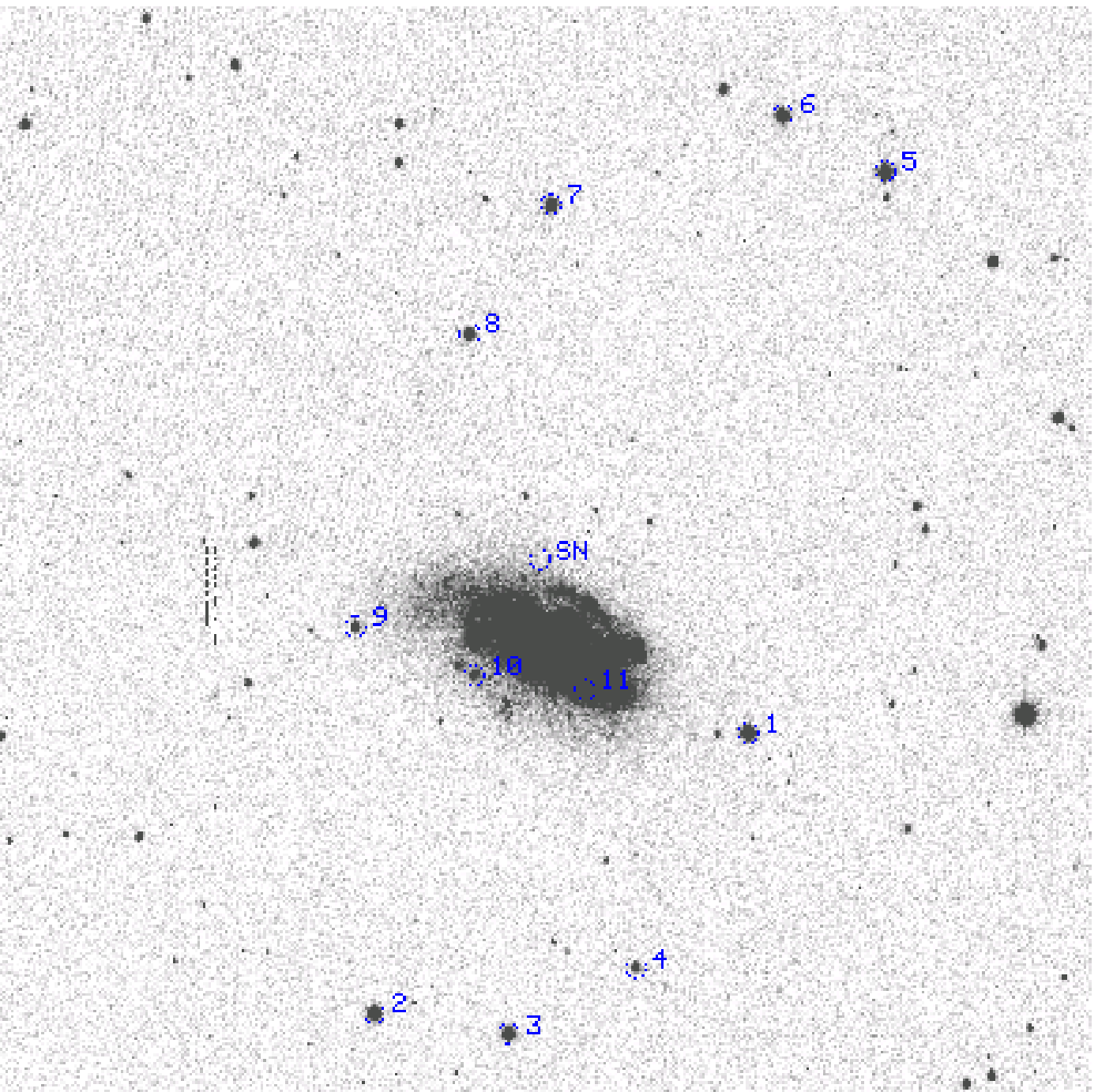]
{NGC 1559, SN~2005df, and the field stars nearby.  This is a 40 second $V$-band
exposure obtained with the CTIO 0.9-m telescope on 23 December 2006 UT.  The
field of view is 10$\farcm$56 by 10$\farcm$56, which is considerably larger
than the ANDICAM field of view. \label{finder}
}

\figcaption[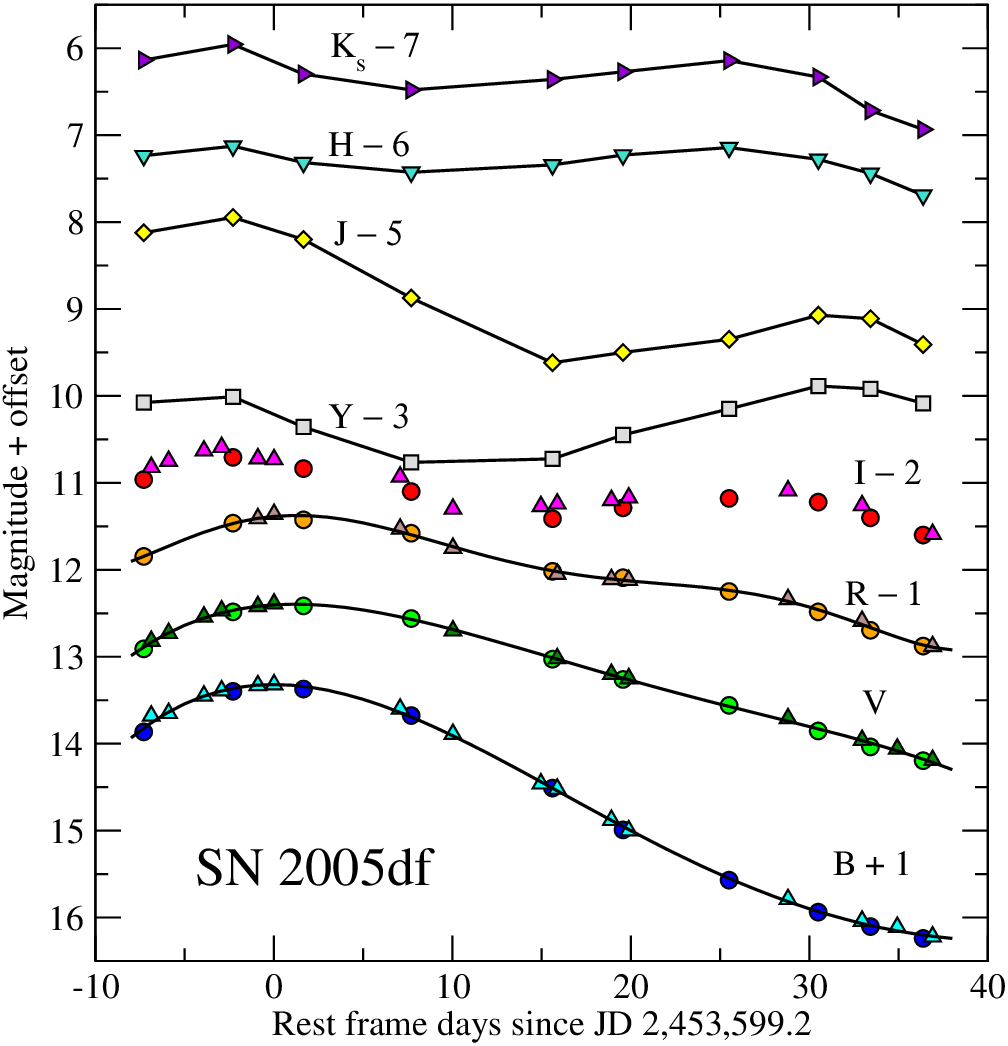] {Optical $BVRI$ and near-infrared $YJHK_s$ light curves
of SN~2005df.  For the optical photometry the colored circles are data from the
CTIO 1.3-m telescope, while the triangles are data from ANU published by
\citet{Mil_etal10}.  All the near-IR photometry is from the CTIO 1.3-m 
telescope.  The CTIO $BV$ photometry has been S-corrected to the system 
of \citet{Bes90}, while the $JHK_s$ photometry has been S-corrected to the
photometric system of \citet{Per_etal98}.  The $BV$ data have been fitted with a
fourth order polynomial, while the $R-$band data have been fitted with a sixth
order polynomial.  The RMS residuals of the $B$-, $V$-, and $R$-band fits are
$\pm$ 0.031, 0.021, and 0.028 mag, respectively.\label{bvri_yjhk}}

\figcaption[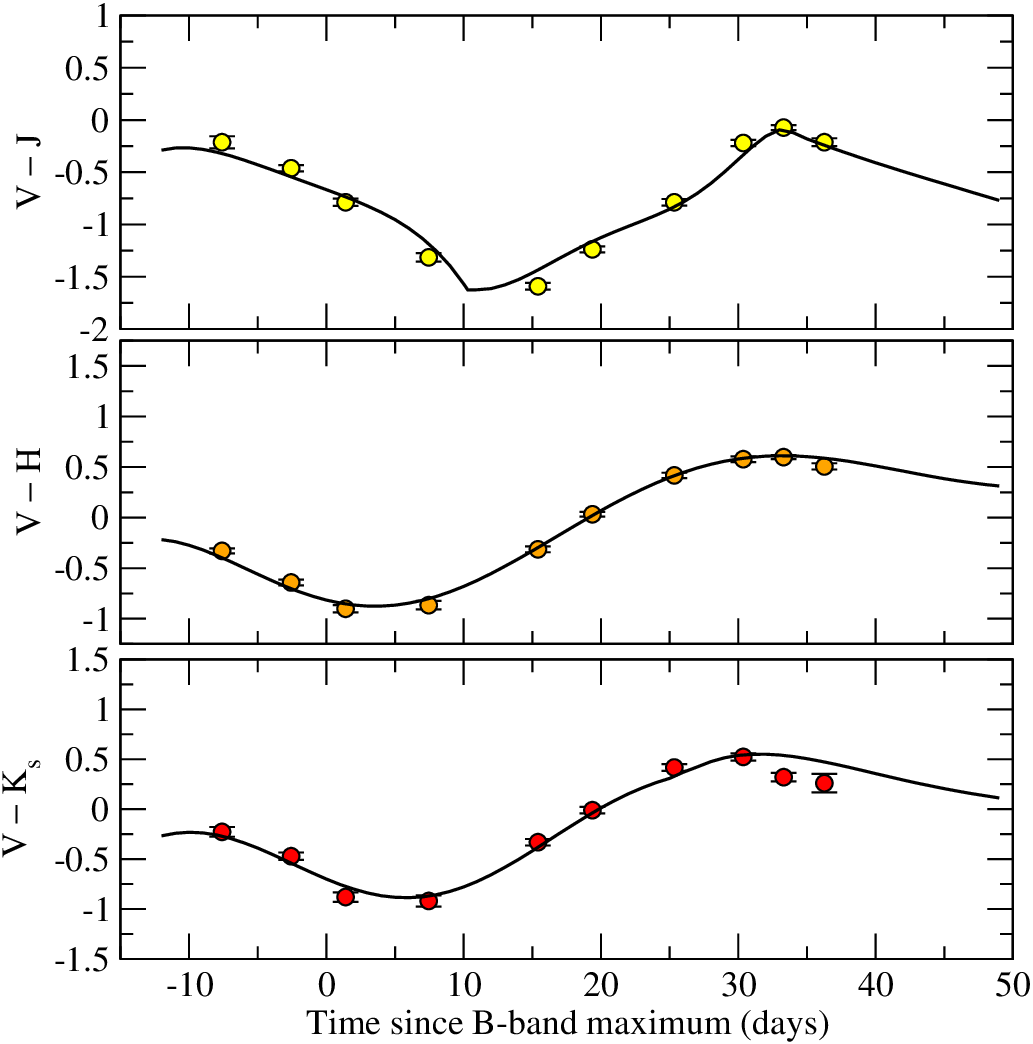] {$V$-[$J,H,K_s$] colors of SN~2005df.  The solid
black lines are from fits to data of SN~2001el.  They have been shifted
in the Y-direction to minimize the reduced $\chi^2$ value of the fits.
\label{vjhk}
}

\clearpage

\begin{figure}
\plotone{df_finder.ps}
{\center Krisciunas {\it et al.} Fig. \ref{finder}}
\end{figure}

\begin{figure}
\plotone{bvri_yjhk.eps}
{\center Krisciunas {\it et al.} Fig. \ref{bvri_yjhk}}
\end{figure}

\begin{figure}
\plotone{vjhk.eps}
{\center Krisciunas {\it et al.} Fig. \ref{vjhk}}
\end{figure}

\end{document}